# Examining effect of architectural adjustment on pedestrian crowd flow at bottleneck

Xiaomeng Shi [a, b, c], Zhirui Ye [a, b, c *], Nirajan Shiwakoti [d], Dounan Tang [e], Junkai Lin [c,]

[a] *Jiangsu Key Laboratory of Urban ITS, Southeast University*

[b] *Jiangsu Province Collaborative Innovation Center of Modern Urban Traffic Technologies*

[c] *School of Transportation, Southeast University*

*Sipailou #2, Nanjing, Jiangsu, 210096, China*

[d] *School of Engineering, RMIT University*

*Carlton, VIC, 3053, Australia*

[e] *Department of Civil and Enviornmental Engineering, Insititute of Transportation Studies,*

*University of California, Berkeley*

*Berkeley, CA, 94720–5800, USA*

**Abstract:** Recent advances in bottleneck studies have highlighted that different architectural adjustments at the exit may reduce the probability of clogging at the exit thereby enhancing the outflow of the individuals. However, those studies are mostly limited to the controlled experiments with non–human organisms or predictions from simulation models. Complementary data with human subjects to test the model's prediction is limited in literature. This study aims to examine the effect of different geometrical layouts at the exit towards the pedestrian flow via controlled laboratory experiments with human participants. The experimental setups involve pedestrian flow through 14 different geometrical configurations that include different exit locations and obstacles near exit under normal and slow running conditions. It was found that corner exit performed better than middle exit under same obstacle condition. Further, it was observed that the effectiveness of obstacle is sensitive to its size and distance from the exit. Thus, with careful architectural adjustment within a standard escape area, a substantial increase in outflow under normal and slow running conditions could be achieved. However, it was also observed that placing the obstacle too close to the exit can reduce outflow under both normal and slow running conditions. Moreover, we could not observe "faster–is–slower" effect under slow running condition and instead noticed "faster–is–faster" effect. In addition, the power law fitted headway distribution demonstrated that any architectural configurations that enhanced the outflow have higher exponent value compared to the other configuration that negates the outflow. The findings from this paper demonstrate that there is a scope to adjust the architectural elements to optimize the maximum outflow at the egress point. Further, the output from the experiments can be used to develop and verify mathematical models intended to simulate crowd evacuation.

---

[*] Corresponding author. Tel.: +86–25–8379–0583; fax: +86–25–8379–4102.

E–mail address: yezhirui@seu.edu.cn

# 1. Introduction

Over the past two decades, there has been steady increase of crowd related incidents such as stampedes and evacuation leading to serious injuries and fatalities [1]. Understanding pedestrian crowd dynamics is critical for efficient and safe crowd management. Crowd dynamics is complex at bottlenecks such as narrow corridors, exits and doorways, under congested and emergency situations. For civil, traffic and safety engineering practices, scientific understanding of crowd motion patterns at bottlenecks is important in planning and designing of pedestrian facilities in order to achieve safe, efficient and comfortable walking operations. In addition, rapid and safe egress is also regarded as an essential requirement for escape during emergencies such as natural disasters (e.g. earthquakes, hurricanes) and human–induced disasters (e.g. fire accidents, terrorist attacks). In the crowd dynamics literature, the term "panic" has been frequently used to refer emergency situations [2–5]. Hence, a clarification on the use of the word "panic" is required. In this study, we consider panic or emergency as situations in which pedestrians are impatient (due to high crowd density and short time for egress), and which result in competitive and pushing behavior [2,6,7]. This is different from the meaning of the term that most social scientists have used, i.e., "panic" is restricted to instances when both high emotional arousal and irrational behavior occur [8–10]. In the past, people have displayed competitive and pushing behaviors during emergency evacuation [2,6,7]. However, there have been also instances where people behaved in much calmer way and helped others [6,9,11,12]. The focus on the former issue is more critical as it can create negative consequences such as stampedes.

Scientists have highlighted that crowd accidents such as stampede, trampling and crushing may have a larger potential to occur at bottlenecks due to complex movement patterns and jamming of dense crowds [2,13–15]. Study on complex crowd movements (e.g. counter flows, merging, crossing) at bottlenecks have demonstrated that such movement pattern creates additional delays due to conflicts arising from multi-directional movements [2,13,16–18]. Similarly, researchers have demonstrated that jamming at bottlenecks creates additional delays [2,15,18]. In an emergency situation, such additional delays can make pedestrians impatient which can then trigger the competitive and pushing behavior resulting in inefficient evacuation or can ultimately lead to deadly consequences such as stampedes [2,7,19]. Shi et al. [13] provides the background information on relevant literature that have highlighted that it is essential to gain a thorough knowledge on the mechanisms of crowd at specific bottleneck sections as any external perturbations in a high density complex movement can create shockwave and stop–go condition. These shockwave and stop-go condition can make the pedestrians impatient and pushy resulting in negative consequences such as stampedes [7]. Further, Shi et al. [13] discusses how the bottleneck such as inverse merging T–junction can lead to stampede with a reference to love parade crowd disaster in Germany [20]. Likewise, researchers have demonstrated that clogging and jamming are likely to happen at bottleneck [2]. In another work, researchers have studied the Akashi fireworks display crowd accident in Western Japan in 2001, specifically with reference to crowd accidents that occurred at bottlenecks such as stairway and around the corner of the bridge [14]. The work by Moussaïd et al. [15] highlights that the well–coordinated motion among pedestrians suddenly breaks down, particularly around bottlenecks, resulting in a largely fluctuating and uncontrollable patterns of motion, called crowd turbulence. Such crowd turbulence has resulted stampede in the past.

Previous mathematical prediction and empirical study show that some counter intuitive crowd movements phenomena such as the "faster–is–slower" (FIS) effect [2] often occur at bottlenecks during egress processes, resulting in clogging, queuing and arching at exits thereby impeding the outflow of people. However, recent study on controlled laboratory experiments with selfish and polite participants have noted that instead of FIS effect, "faster–is–faster" (FIF) effect was observed in the experiments [21]. Models also present a very surprising prediction that outflow of people will be enhanced if there is a partial obstruction or column on the "upstream" side of an exit [2]. This counterintuitive performance of the obstacle near an exit has also been the subject of interests for researchers working on granular matters [22,23] and architecture [3,24]. In addition, recent innovative experiments with non–human organisms have highlighted that the minor adjustments of architectural configurations at bottleneck can have large effect in terms of outflow of the individuals similar to those mathematical predictions [4,25–27]. However, these predictions have not been fully tested with empirical data on humans. The existing empirical studies with human subjects at exit point have mostly looked into the effect of different exit widths on the outflow of individuals through controlled laboratory experiments.

In summary, to our knowledge, there has been no systematic and comprehensive study on empirical data with human subjects that has examined the performance of different architectural adjustments at the bottlenecks. Therefore, the understanding of the performance of different architectural configurations at the bottleneck demands specific attention. This study aims to examine the performance of different architectural adjustments within a given escape area on the outflow of pedestrians via controlled laboratory experiments with pedestrians. Hence, this empirical study with human subjects will provide valuable insights to supplement and extend the knowledge towards the effect of architectural adjustments to optimize the maximum outflow through the egress point as well as in verification of the existing evacuation modeling and simulations.

The paper is organized as follows. The next section summarizes the related literature on crowd movement through bottlenecks. It is then followed by the description of our methodology including data collection through controlled laboratory experiments. This is followed up with the data analysis and discussions. The final section presents the conclusions and recommendations for future research.

## 2. Literature review

This section is structured into two parts. The first subsection reviews the simulation models that have been adopted to simulate crowd egress dynamics at bottlenecks. Then, the second subsection describes the empirical studies on bottleneck flows derived from biological entities including human participants and non–human organisms through controlled experiments.

*2.1 Simulations of crowd egress at bottlenecks*

In the past, simulation studies were conducted to reproduce crowd behaviors at bottleneck [2–4,17,27–37] and examine performance of architectural adjustments near a bottleneck or an exit [3,5,28,29,37–40]. Helbing et al. [2] simulated the crowd egress behavior at bottleneck (exit) under panic situation and noted FIS effect due to impatience of pedestrians. Further, through the simulation, they noted that FIS effect can be avoided by improved outflows through suitably placed columns in front of the exits, which also prevent the buildup of fatal pressures. Using the social force model developed by Helbing et al. [2], Escobar and De La Rosa [3] examined several architectural adjustments including obstacles (columns) placement near an exit on the outflow of panicked crowd. They found that in general, obstacles slow down the individual velocity and increase the outflow. In another study, Frank and Dorso [28] simulated two types of pedestrians, "strategic–managing" and "non–strategic" to examine the effect of obstacles placement near an exit. The "strategic–managing" pedestrians would change his (her) desired direction in order to avoid obstacles in the way out. In contrast, the "non–strategic" pedestrian would move directly to the exit, as if the blocking obstacles were transparent. The researchers found that reduction in evacuation time dependent on the distance of the obstacle from the exit. For example, When the obstacle was very close to the exit, evacuation time was reduced for both, "non–strategic" and "strategic pedestrians", with respect to the obstacle–free situation. However, when the obstacle was far away from the exit, the evacuation time for "non–strategic" and "strategic pedestrians" was similar to the obstacle–free situation. Likewise, Shiwakoti et al. [39] simulated the pedestrians escape through an exit with/without columns as well as the effect of location of exit (corner vs. middle exit). It was found that, in general, the placement of column near an exit increases the outflow of the pedestrians. However, when the obstacle size and location of the column were varied, the outflow was less than the column free situation in some cases. These differing decisions about the performance of the architectural adjustments in an escape area have been noted by other researchers as well [4,5,17,27,29–37]. In another study, Parisi and Patterson [40] adopted two simulation models to examine the effect of the length of the bottleneck after the exit and the distance of the exit to the lateral wall, on evacuation time. Result suggested that in order to minimize the evacuation time, bottleneck length should be reduced to zero if possible. Meanwhile, two simulation models provided different results on the performance of the exit closer to the lateral wall. Hence, it was highlighted that the collection and analysis of empirical data that considers different geometries is critical to improve our understanding on how the exit position influences the evacuation time. In a recent study by Zhao et al. [5], it was demonstrated through numerical modeling that the obstacle near exit creates a significant reduction of high density region by effective separation in space which then leads to the increase of escape speed and evacuation outflow. The study further highlighted the need of human experiments to verify the model's prediction and enhance our understanding on the effectiveness of obstacles to reduce the evacuation time.

The obstacles effect near an exit has been of interests not only to crowd researchers, but also researchers working on granular flow [22,23]. For example, inspired by previous simulation results of pedestrian crowd dynamics on the performance of obstacles near an exit, Zuriguel et al. [23] examined the effect of inserting an obstacle just above the outlet of the silo on the clogging process. They found that if the position of the obstacle is properly selected, a substantial decrease of the clogging probability can be obtained. It was suggested that the obstacle induces a strong reduction of the pressure above the outlet. The researchers suggested that this behavior could have an analogy in the flow of crowds through bottlenecks, where it has been typically assumed that the main role of a column behind an exit is to prevent straight motion of people towards the exit. Further, Alonso–Marroquin et al. [22] concluded through the particle–based simulation that the flow rate in a hopper with an obstacle placed before the bottleneck was dependent primarily on the aperture (the minimum distance between the obstacle and the hopper) and the angle of the hopper. Contrary to expectations, researchers found that the flow rate across a bottleneck actually increases if an optimized obstacle is placed before it. These finding emphasizes the need to further explore the role of obstacle size and position from the exit towards improved outflow of particles or individuals, especially with empirical data.

*2.2 Empirical data of egress dynamics at bottlenecks*

It is critical to have empirical data to test the model's predictions. As such, there have been several controlled laboratory experiments to study the egress dynamics at bottlenecks [21,41–49], but less empirical study on the performance of architectural adjustments or obstacle effects near an exit [4,17,27,39,50,51]. For ethical and safety reasons, experiments with human participants at bottlenecks were restricted to non–panic situations including normal condition and orderly evacuation drills. To overcome the scarcity of data under panic conditions, experiments with non–human organisms have been conducted under stressed [27,50] or panic conditions [4,33,39,52,53], especially to examine the effect of architectural features or obstacles effect near the exit.

In line with the simulation studies, most empirical studies paid attentions on the bottleneck width $b$ as it was regarded as one of the most significant influence factors on capacity $J_s$. In earlier studies, the estimated relations between $b$ and $J_s$ have been included in some well–known handbooks and guidelines and the comparison statistic has been presented in [43]. Then, many controlled laboratory experiments with human participants were performed to test the specific flow rate with different bottleneck widths [17,41,42,44,45]. Later, more variables were considered in the experiments such as populations, bottleneck geometrics, door open conditions, light intensities, stress levels, competitiveness and politeness [21,49,54,55].

However, seldom of these empirical studies with humans systematically looked into the effect of architectural adjustments (e.g. obstacles, location of exits) towards outflow near a bottleneck [17,51].

Given the lack of empirical data under panic situations, egress at bottleneck were studied through experiments using non–human organisms such as ants [4,39,56,57], mice [33,52] and sheep [58] under panic or stress condition. Saloma et al. [33] performed an experiment with panicked mice escaping through doors of various widths and separation. It was observed that the mice displayed self–organized queuing in an order when the exit was only large enough for a single mouse. As the width of the exit increased, the queuing disappeared and the mice displayed competitive behavior resulting in blockage and making their escape inefficient. Thus, the mice escaped via an exit in bursts of different sizes that obeyed exponential and power–law distributions (truncated) depending on exit width. Further, the researchers noted that over–sampling or under– sampling the mouse escape rate prevented the observation of the predicted features and as such they highlighted that it is important to establish the critical sampling rate to reveal the true escape dynamics of the biological entities. Recently, Lin et al. [59] conducted experiments with mice passing through an exit with or without column under high competition. The study revealed that the presence of an obstacle in front of an exit can improve or reduce the evacuation efficiency depending on the size and location of the obstacle.

Moreover, Garcimartín et al. [58] observed the competitive behavior of a sheep herd passing through a narrow door and evaluated the effect of increasing the door size and the performance of an obstacle placed in front of the door. The authors highlighted that the action of widening the exit or placing an obstacle before the door may have a beneficial effect in terms of reducing the clogging probability. The authors recommend that the effectiveness of those architectural adjustments can be gauged by measuring the exponent of the power law tail in the time lapse distribution. And recently, Zuriguel et al. [50] further extended Garcimartín et al. [58]'s results by considering three different obstacle positions (60, 80, 100 cm from the door). They found that the sheep flow could be improved in the 80 cm and 100 cm cases while worsened in the 60 cm case. But they did not consider the effects of obstacle size and exit position in this study.

Regarding ants experiments, Shiwakoti et al. [4] found that the placement of partial obstruction like a column near an exit enhances the outflow of ants as compared to the scenario when there was no column near an exit under panic conditions. Further, Shiwakoti et al. [56] conducted experiments with ants and found that the placement of exit at corner is more efficient than the middle exit in a given escape area. Those observations led the researchers to conclude that perhaps the effect of adjustment of small architectural features may have substantial effect in the outflow of individuals of varying body sizes [60] though the underlying governing mechanisms of escape behavior could be different among those entities. Later, Soria et al. [57] conducted ants egress experiments and observed the FIS effect among escaping ants. However, they concluded that the FIS effect was not generated by the occurrence of blocking clusters right before the exit as in simulation of human crowds because ants did not display a selfish evacuation behavior. Parisi et al. [61] revisited the ants experiments in Soria et al. [57] and found that the FIS effect among ants could not be reproduced by the simulation model intended for human crowd simulation. They also confirmed the findings of Soria et al. [57] that FIS effect in ants was not caused by selfish behavior.

In summary, none of these empirical studies have considered the impact of both exit position and obstacle placements in a systematic way. Further, those experiments conducted with non–human organisms (with/without obstacle, corner/middle exit) have not been tested with human empirical data. Although simulation models have examined the effect of obstacle placements in more detail, they lack the verification with empirical data of human subjects. Therefore, a detailed and systematic empirical study with human participants on understanding the effect of architectural adjustments towards the outflow of pedestrians as conducted in this study will bridge the gap between engineering, granular flow and biology.

## 3. Methodology
### 3.1 Controlled experiments with human participants
#### 3.1.1 Experimental setup

The primary objective of conducting the experiments is to examine the impact of different architectural configurations on crowd egress through narrow bottleneck in a given area. The design of experimental setup was inspired from the previous studies with ants by Shiwakoti et al. [56] which examined the performance of egress of ants with/without obstacle near an exit and corner vs. middle exit within the same escape area. As such, four controlled variables on the exit designs were selected as below:

- Exit position: middle exit / corner exit
- Obstacle placement: with / without obstacle
- Column size – the diameter of column shaped obstacle $\phi$ : 60 cm / 100 cm

- Column position – the distance from column's outer edge to the midpoint of exit line $D$: 60 cm / 80 cm / 100 cm

The shape, size and position of column are designed in line with the results from previous simulation studies on obstacle effects [3,17,22,28,29,39].

According to the orthogonal design method, a total of 14 sets of experiment scenarios were established and the scenarios were illustrated in Table 1. In addition to the variables of design elements, desired speed levels, which are also regarded as an important indicator of competitiveness [62], are also considered in our experiments as below:

- Desired speed levels: normal walking speed / slow running speed

Normal walking is often relevant to the normal situations such as the daily pedestrian activities or special gathering events. Meanwhile, slow running (or faster walking) can be more representative when people are in hurry (as observed in train stations) or in evacuation process [16]. Because of ethical and safety reasons, it is not feasible to perform real panic experiments in which participants' irrational behaviors such as hurting themselves and others are allowed [8]. Therefore, in our experiments, all the participants were in a rational status and without any cognitive deficits. Dangerous behavior such as pushing was not allowed during the egress processes. But we did not control the normal aggressiveness of the participants so that pedestrians could display politeness or selfishness behavior during the experiments.

Although higher number of repetitions may be desirable for statistical analysis, considering resource and cost constraints, three to five repetitions have been sufficient to conduct relevant analysis for laboratory walking experiments [18]. In addition, with greater number of repetitions, there is a risk that participants may exhibit learned behavior which may not be realistic unless different set of participants are used in the experiment. This can however increase the costs and resources required for the experiments. We further tried to minimize the cumulative learning behavior of participants by introducing the randomness in their location with each repetition. Finally, a total of 14×2×3=84 runs were performed. The sequences of the experiments were based on the experiment number as shown in Table 1.

**Table 1**
Orthogonal design table for different experiment setups

| Exp. No. | Exit position | | Column size (cm) | | | Column position (cm) | | |
|---|---|---|---|---|---|---|---|---|
| | Middle | Corner | 0 | 60 | 100 | 60 | 80 | 100 |
| 1 | • | | • | | | | | |
| 2 | • | | | • | | • | | |
| 3 | • | | | • | | | • | |
| 4 | • | | | • | | | | • |
| 5 | • | | | | • | • | | |
| 6 | • | | | | • | | • | |
| 7 | • | | | | • | | | • |
| 8 | | • | • | | | | | |
| 9 | | • | | • | | • | | |
| 10 | | • | | • | | | • | |
| 11 | | • | | • | | | | • |

| 12 | • | | • | • | | |
| 13 | • | | • | | • | |
| 14 | • | | • | | | • |

Note: "•" represents the selected parameter

The series of experiments were conducted in an indoor basketball training stadium located at a college in Suqian, China, during May, 2016. A square chamber with a dimension of 8m×8m×1.8m (length×width×height) was built as the experiment area. The detailed dimension of the room is shown in Fig. 1. A fixed door width 1.2 m was adopted in our experiments which could enable two individuals to pass the door at the same time. A waiting line was placed at 2 m behind the exit and the region behind the waiting line was considered as waiting area.

The boundary was made of opaque dark green material and the 1.8 m height could enable a blocked vision condition as the maximum height of participants' apparent horizon was lower than 1.8 m. Further, the stability of the exit side boundary was enhanced by steel girders so that they remain stable when participants exited the chamber. The obstacles used in the experiments were cylinder shaped columns made of metal material over 100 kg in weight.

Two stereo–cameras with a resolution of 1920×1080 were implemented on the roof of the stadium (about 9 m from the ground surface). Cam. #1 was a local camera placed perpendicular to the ground above the exit region as shown in the blue–shaded area shown in Fig. 1. (c). Cam #2 was a global camera placed on the far side of the stadium so that it could capture the full vision of the experiment field. Also, an unmanned aircraft vehicle (UAV) was adopted to take some snapshots during the experiments as shown in Fig. 2. (d).

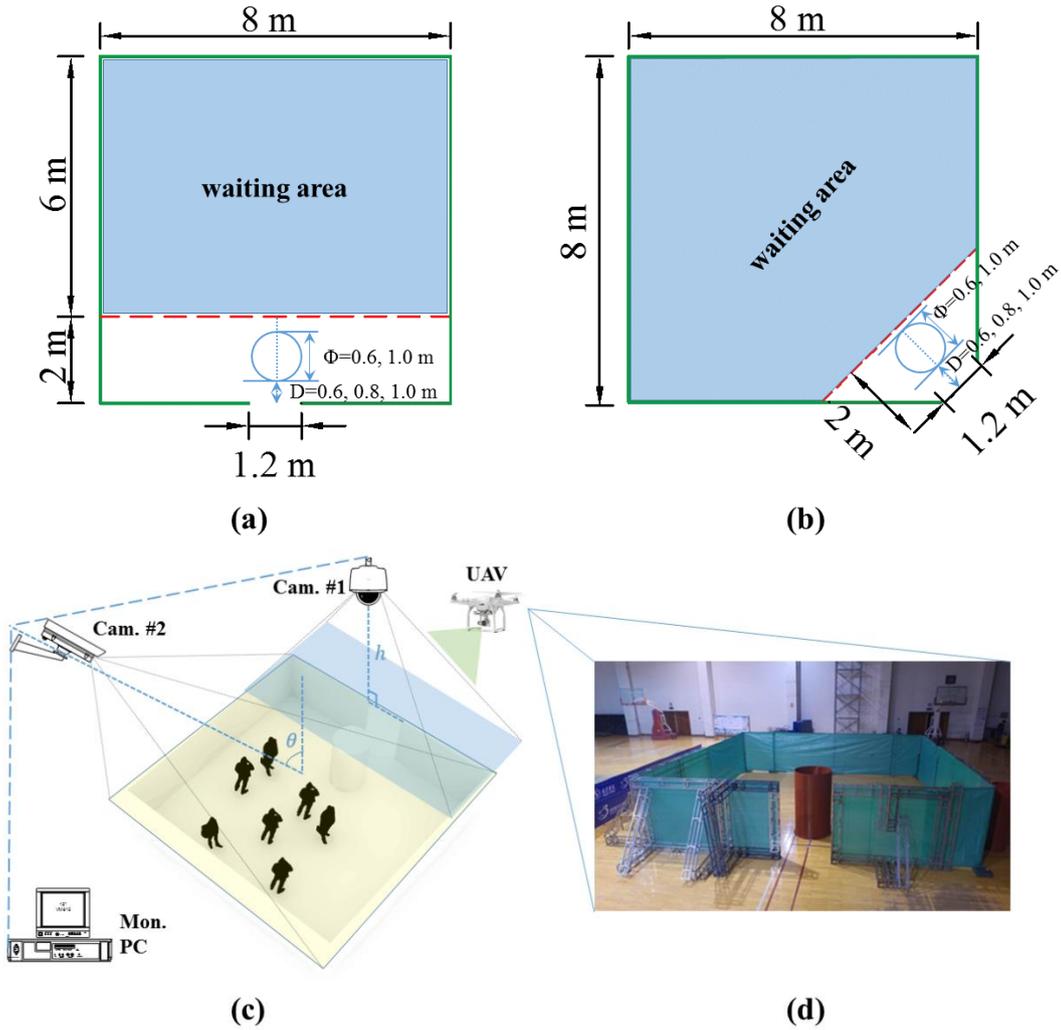

**Fig. 1.** Experiment setups for the human crowd egress experiments: (a) middle exit graphic designs; (b) corner exit graphic designs; where $\phi$ represents the diameter of column, and $D$ refers to the distance from column's outer edge to the midpoint of exit line. (c) 3D conceptual diagram of cameras positions and video data transmission process; (d) an overview snapshot of experiment site recorded by a UAV.

*3.1.2 Participants*

A total of 50 participants (26 males and 24 females) were involved in this series of experiments. The number of participants considered in this experimental setup was enough to create Level of Service E (>0.7–1.4 m$^2$/pedestrian) [63]. They were volunteer students from different departments and year levels of the local college. This would, to some extent, ensure the unfamiliarity among participants that would be expected in a real world situation, where pedestrians in a crowd may not know each other. Each participant was paid 50 RMB along with a free lunch ticket (about $10 cost for each individual a day). Also, five coaches from the local college were also invited to assist in the organization of the experiments.

To increase the robustness of automatic data extraction from videos, participants were asked to wear a uniform white T–shirt and a color cap (pink for female, orange for male) during the entire experiment process. For manual validation purpose, their shoulders and heads were tagged with labels marked with numbers from 1 to 50.

Before the experiments, all the participants were asked to fill a form about the required information on their personal attributes such as gender, age, weight, height, body width (shoulder width) and body length (body thickness). The statistical results (mean value, standard deviation and maximum value) are shown in Table 2.

Moreover, some important psychological properties (such as desired free flow speed, reaction time and shy away distance) were investigated and the statistics are also presented in Table 2. These properties have been frequently adopted by many simulation models and could significantly affect pedestrian walking behaviors [42,64]. The desired free flow speeds for each participant were measured by asking them to walk/run freely in a 30m–length straight corridor for three repetitions. The desired speed that we measured refers to the free flow speed of individual walking which could be useful to the researchers who may use to calibrate the crowd simulation models. However, it is to be noted that the desired velocity of each individual could change under different walking environment and their enthusiasms in participating the experiments may decrease due to the tiredness if many repetitions are conducted. As such, we conducted only three repetitions and provided enough rest time between different experimental setups. We did not notice any tiredness among participants during experiments. By recording the inlet time $t_{inlet}$ and outlet time $t_{outlet}$ the mean free flow speed $v_0$ for each individual was calculated: $v_0 = 30/(t_{outlet} - t_{inlet})$. Then, the statistics of desired speed levels for all 50 participants were obtained. Reaction time $t_{react}$ was defined as the time interval between the time point when an individual received a signal to walk/run $t_{signal}$ to his/her actual movement time $t_{move}$, $t_{react} = t_{signal} - t_{move}$. Shy away distance was estimated by the average lateral space when each individual walk/run in the corridor [65].

**Table 2**
Statistics for the individual attributes of participants

| Personal properties (unit) | Mean value | Standard deviation | Min–Max value |
|---|---|---|---|
| Age (years) | 21 | ±2.5 | 19–25 |
| Weight (kg) | 62 | ±11 | 43–95 |
| Height (cm) | 168 | ±7 | 153–183 |
| Body width (cm) | 42 | ±6 | 32–58 |
| Body length (cm) | 36 | ±11 | 30–50 |
| Desired speed for normal walking (m/s) | 1.31 | ±0.13 | 1.09–1.78 |
| Desired speed for slow running (m/s) | 2.62 | ±0.26 | 2.02–3.22 |
| Reaction time (s) | 0.62 | ±0.68 | 0.21-2.56 |
| Shy away distance (cm) | 13 | ±5 | 9–25 |

*3.1.3 Experimental procedure*

The experiment was conducted for an entire day. Experiment sets #1 to #7 were performed in the morning and #8 to #14 in the afternoon. Participants were asked to stand in the waiting area of the room. Before the start of each repetition, the spatial distribution pattern of the participants was adjusted randomly by the coaches and participants were only informed regarding their desired speed levels to be adopted and their destinations. However, no information was provided to the participants regarding the research aims of the study, so that they walk normally (as in their daily life). Upon hearing the whistling signal, all the participants started to move towards the exit and exited the room. After completing each three repetitions, participants were allowed to rest for about 10–15 minutes. Some snapshots from Cam. #1 during the experiments are shown in Fig. 2.

### 3.1.4 Data extraction

Trajectories of pedestrians were automatically extracted from the recorded video footages using a pedestrian multiple tracking software PeTrack developed by [66]. The original video sequences were 25 frames–per–second (fps). Some examples of the extracted trajectories were shown in Fig. 2. Line time of each individual passing the exit was measured automatically and the time headway could be derived using Equation 1 (presented in next section).

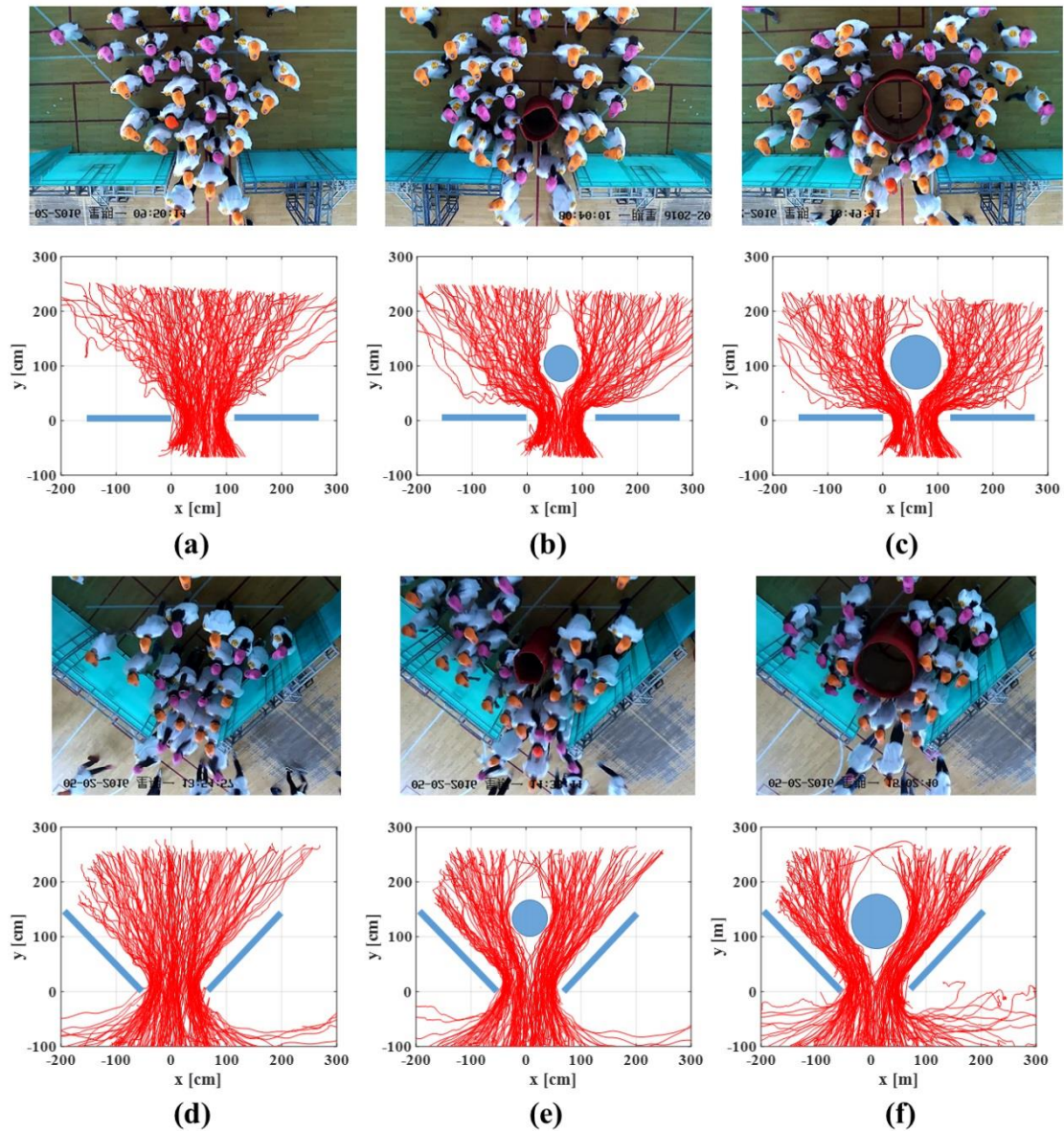

**Fig. 2.** Snapshots and trajectories examples of the experiments: (a) 1. Middle exit (standard design); (b) 3. Middle exit, $\phi$=60 cm, $D$=80 cm; (c) 5. Middle exit, $\phi$=100 cm, $D$=60 cm; (d) 8. Corner exit; (e) 11. Corner exit, $\phi$=60 cm, $D$=100 cm; (f) 13. Corner exit, $\phi$=100 cm, $D$=80 cm. The snapshots were selected from one frame during one repetition and the trajectories were the collection of all the three repetitions under one kind of desired speed level. (a), (b), (c) represent normal walking situation, while (d), (e), (f) illustrate slow running condition.

*3.2 The measures for egress dynamics at exits*

*3.2.1 Total evacuation time and mean time headway*

Traditionally, total evacuation time (TET) along with mean time headway ($<\Delta t>$) are the most widely used approach to evaluate evacuation efficiency as shown in Equation 1. The exit time of each individual $t_i$ is usually recorded by placing a detector at the measuring segment.

$$TET = t_N - t_1$$
$$\Delta t = t_{i+1} - t_i \quad (1)$$
$$<\Delta t> = (t_N - t_1)/N$$

where $t_1, t_N$ represent the time of the first and last person passing the exit, respectively; $N$ denotes the number of pedestrians passing the measuring segment.

*3.2.2 Definition of flow*

We adopt a traditional approach to calculate the flow of pedestrians $J$ (Ped/s) passing an exit i.e. the number of pedestrians past the exit segment within a unit of time:

$$J = \frac{N}{t_N - t_1} = \frac{N}{\sum_{i=1}^{N}(t_{i+1} - t_i)} = \frac{1}{<\Delta t>} \quad (2)$$

where $<\Delta t>$ refers to the mean time headway for all passing individuals.

Given the width of the exit $b$, the specific flow rate $J_s$ (Ped/s/m) is derived:

$$J_s = \frac{J}{b} \quad (3)$$

Select a standard design $J_{s_0}$, the relative efficiency for specific flow rate $RE_{J_s}$ is calculated:

$$RE_{J_s} = \frac{J_{s_i} - J_{s_0}}{J_{s_0}} \times 100\% \quad (4)$$

**4. Data analysis**

*4.1 Exit time and flow*

To present straightforward results of crowd escape order, we plotted the escape order i.e. the exit time of each individual as shown in Fig. 3, (a) for normal walking and (b) for slow running. For each set of experiment, the values of exit time were averaged over individuals for all the repetitions.

We could directly observe the trend of flow and clogging during the egress process from a temporal evolution of the escape order plots as shown in Fig. 3 (empirical data on human crowd for both normal and slow running situations). Steeper ascent of the plot indicated higher flow, smoother ascent meant lower flow and flat ascent depicted clogs. From these figures, we could compare the general trend of different outcomes among different experiments. For example, in terms of exit position, corner exits tend to have higher efficiency in terms of outflow as compared with middle exit under the same obstacle condition. The placement of obstacle did not always increase the outflows, it depended on the obstacle size and distance to the exit.

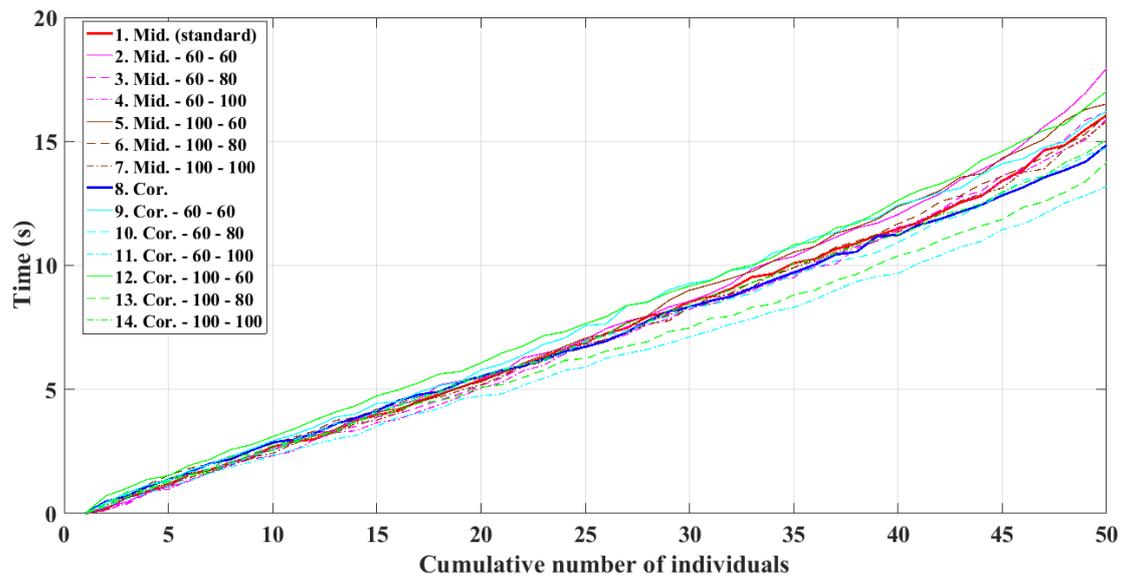

(a)

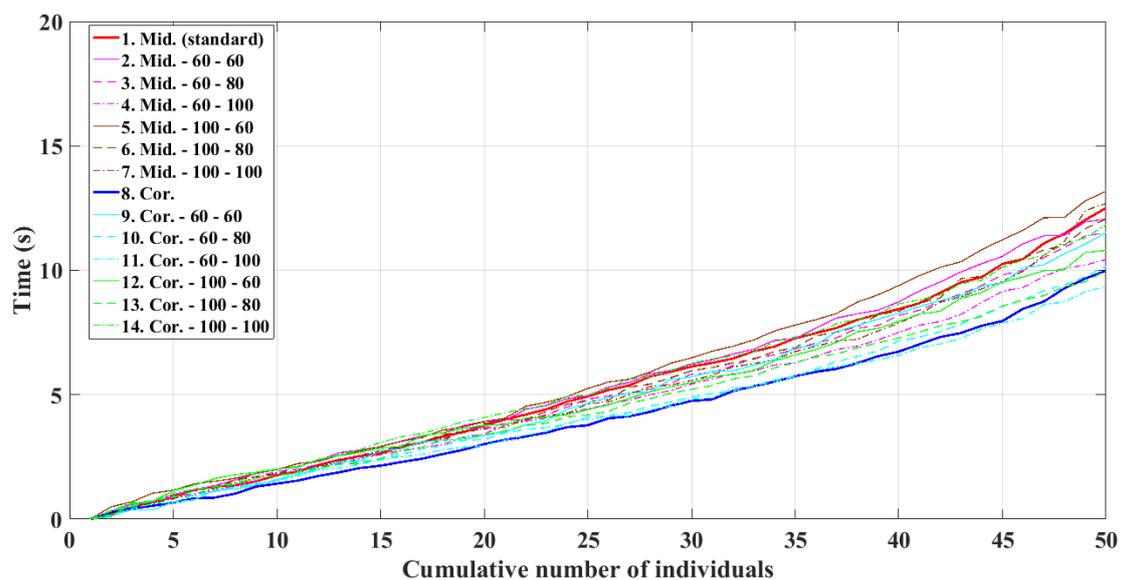

(b)

**Fig. 3.** Escape order plots of empirical data averaged over individuals of all the repetitions under (a): normal walking condition; (b): slow running condition

Moreover, the specific flow rates $J_s$ for all scenarios of all repetitions were calculated using Equation 2 & 3. The mean values of $J_s$ with standard deviation (S. D.) were listed in Table 3. Selecting #1 middle exit without obstacle as the standard design, the relative efficiencies (R. E.) of $J_s$ were also calculated and displayed in Table 3.

**Table 3** Mean value and standard deviation (S. D.) of average time headway $<\Delta t>$ (s) and specific flow rate $J_s$ (Ped/s/m) and relative efficiency (R. E.) (%) of $J_s$ (select #1 as standard design) for all human experiment sets under normal walking and slow running situations.

| Exp. No. | Normal walking situation | | | | | Slow running situation | | | | |
|---|---|---|---|---|---|---|---|---|---|---|
| | $<\Delta t>$ | S.D. | $<J_s>$ | S.D. | R.E. | $<\Delta t>$ | S.D. | $<J_s>$ | S.D. | R.E. |
| 1 | 0.32 | 0.06 | 2.67 | 0.55 | – | 0.25 | 0.02 | 3.37 | 0.22 | – |
| 2 | 0.35 | 0.02 | 2.38 | 0.12 | -10.6 | 0.25 | 0.01 | 3.37 | 0.18 | -0.1 |
| 3 | 0.32 | 0.02 | 2.58 | 0.17 | -3.2 | 0.23 | 0.01 | 3.62 | 0.16 | 7.5 |
| 4 | 0.32 | 0.02 | 2.63 | 0.20 | -1.4 | 0.22 | 0.02 | 3.84 | 0.42 | 14.0 |
| 5 | 0.34 | 0.03 | 2.48 | 0.22 | -7.1 | 0.26 | 0.01 | 3.18 | 0.08 | -5.6 |
| 6 | 0.32 | 0.01 | 2.59 | 0.07 | -2.8 | 0.24 | 0.03 | 3.45 | 0.35 | 2.3 |
| 7 | 0.32 | 0.02 | 2.64 | 0.15 | -0.9 | 0.25 | 0.08 | 3.48 | 0.95 | 3.3 |
| 8 | 0.30 | 0.01 | 2.81 | 0.08 | 5.3 | 0.20 | 0.01 | 4.18 | 0.27 | 24.2 |
| 9 | 0.32 | 0.01 | 2.57 | 0.09 | -3.6 | 0.23 | 0.02 | 3.64 | 0.35 | 8.0 |
| 10 | 0.29 | 0.00 | 2.83 | 0.03 | 6.0 | 0.20 | 0.02 | 4.12 | 0.41 | 22.3 |
| 11 | 0.26 | 0.03 | 3.18 | 0.34 | 19.4 | 0.19 | 0.01 | 4.47 | 0.19 | 32.7 |
| 12 | 0.34 | 0.01 | 2.45 | 0.11 | -8.1 | 0.22 | 0.01 | 3.86 | 0.12 | 14.6 |
| 13 | 0.28 | 0.01 | 2.95 | 0.12 | 10.5 | 0.20 | 0.01 | 4.19 | 0.22 | 24.2 |
| 14 | 0.30 | 0.01 | 2.77 | 0.10 | 3.8 | 0.24 | 0.05 | 3.62 | 0.70 | 7.4 |

From the data in Table 3, we could conclude that for normal walking situation, comparing with standard design ($<J_s>$=2.67 Ped/s/m, S.D =±0.55 Ped/s/m), the most efficient setup is #11 (Corner exit, $\phi$=60 cm, $D$=100 cm) with 19.4% R. E. ($<J_s>$=3.18 Ped/s/m, S.D =±0.34 Ped/s/m), and the least efficient geometry is #2 (Middle exit, $\phi$=60 cm, $D$=60 cm) with –10.6% R. E. ($<J_s>$=2.38 Ped/s/m, S.D =±0.12 Ped/s/m). For slow running condition that is more representative of emergency evacuation condition, comparing with standard design ($<J_s>$=3.37 Ped/s/m, S.D =± 0.22 Ped/s/m), the most efficient setup was also #11 with 32.7% R. E. ($<J_s>$=4.47 Ped/s/m, S.D =±0.19 Ped/s/m), and the least one turned into #5 (Middle exit, $\phi$=100 cm, $D$=60 cm) with –5.6% R. E. ($<J_s>$=3.18 Ped/s/m, S.D =±0.08 Ped/s/m).

In general, corner exits performed better than middle exits (lower mean time headway, higher specific flow rate and relative efficiency) under the same obstacle condition. One–way repeated measures ANOVA tests were performed to examine the relationship between exit position and mean time headway ($<\Delta t>$) without obstacles (Exp. #1 vs. #8) for both normal ($F$=13.2>$F_{crit}$, $p<0.01$) and slow running conditions ($F$=8.48>$F_{crit}$, $p<0.05$). And the results showed that exit position had a significant influence on the value of $<\Delta t>$ under both desired speed levels.

Further, obstacles were more beneficial to corner exits than middle exits. Also, obstacles had better performance under slow running situation than normal condition. The results also indicated that obstacle size and distance to the exit might have an influence on the performance of crowd egress from the fluctuations of the data. To examine the statistical relationships between obstacle condition (i.e. column size, column position) and the value of $<\Delta t>$, two–way repeated measures ANOVAs were conducted. We divided the factors into four categories and performed the ANOVAs respectively and the results were as follows: middle exit normal walking ($F=1.65<F_{crit}, p=0.23$), corner exit normal walking ($F=7.39>F_{crit}, p<0.01$), middle exit slow running ($F=3.81>F_{crit}, p<0.01$) and corner exit slow running ($F=21.76>F_{crit}, p<0.01$). From the results of ANOVAs, we could see that the ANOVA test only failed under the condition of middle exit and normal walking where obstacle size $\phi$ and distance to the exit $D$ did not have significant impact on the value of $<\Delta t>$. In other categories, the ANOVAs test passed i.e. $\phi$ and $D$ have a significant influence on $<\Delta t>$ value.

Moreover, by comparing the $<J_s>$ between normal walking and slow running, we could conclude that the specific flow for slow running was always larger than normal walking. Therefore, in our experiments, FIS effect was not observed. This finding was in contrast with Garcimartín et al. [49] but in align with FIF effect with Nicolas et al. [21]. We acknowledge that FIS effect could be more prevalent during strong competitive behavior (for e.g. pushing) and higher degree of threat (for e.g. real fire evacuation vs. evacuation drill) and as such needs further verification of FIS effect and FIF effect. However, it seems under normal and moderate aggressive behavior, FIF effect may be prevalent. Also, such discrepancy may arise from the different architectural configurations, exit width, density of people etc.

### 4.2 Time headway distribution

In order to examine the flow fluctuations during egress process, we introduced the temporal distributions of two consecutive passage of pedestrians (time headway). Time headways were calculated from the exit time of each individual using Equation 2. We displayed the distribution of time headways using the box–and–whiskers plots (boxplots) as shown in Fig. 4. We used the entire sets of data collected in this paper: empirical data for human crowds derived from all the 14 experiment setups (with the aggregation of all the repetitions) under normal and slow running conditions.

From the boxplots, we could observe the quartiles, the whiskers and the outliers of the time headway data. From the bottom to the upper of the vertical boxplots, each transverse line represented the lower whisker, lower quartile (Q1, 25th), median (50th), upper quartile (Q3, 75th), upper whisker, respectively, and the outliers (marked with red '+' individual dots).

The general distribution of $\Delta t$ could be reflected by the position of median line in the boxplots and the overall position of the entire boxplots. In line with the above analysis on flow, the more shorter $\Delta t$ and less longer $\Delta t$ could reflect a better evacuation performance. As noted from Fig. 4 (a) and (b), as the growth of desired speed, the degree of dispersion and skewness in $\Delta t$ increased, while the median (as well as the position of the whole box) decreased slightly.

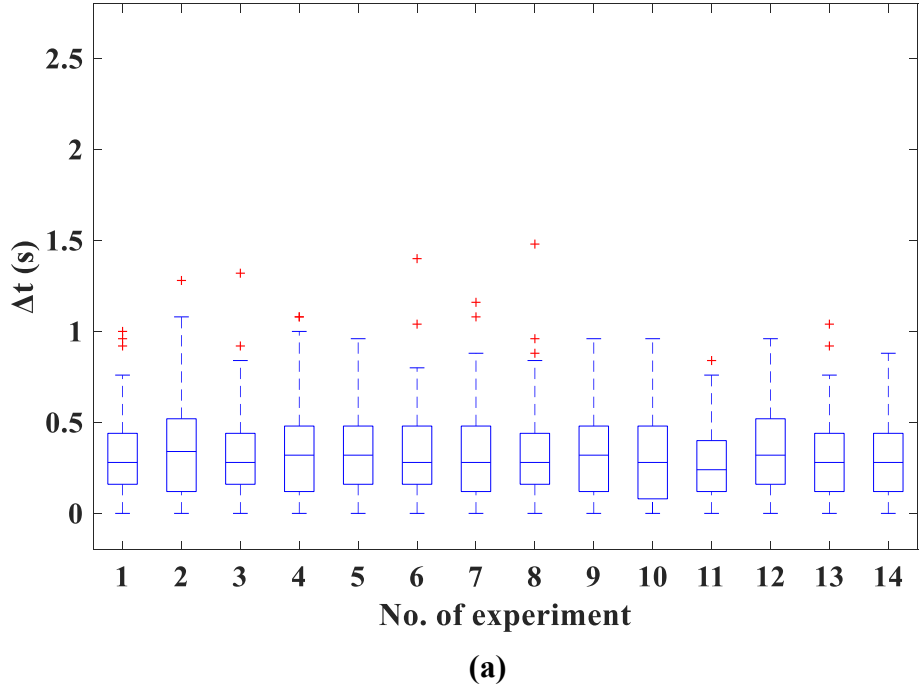

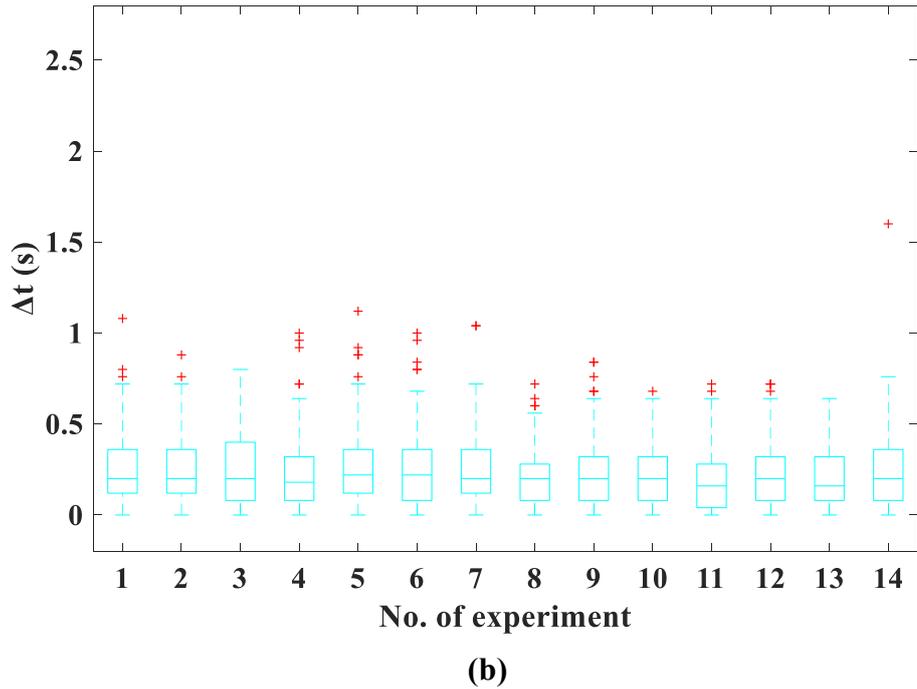

**Fig. 4.** Boxplots of time headways $\Delta t$ of human experiments (a): for normal walking; (b): for slow running

*4.3 Power–law analysis of time headway*

The distributions of time headways have been mostly fitted to different laws such as log–normal, exponent and stretched–exponent [16,65,67,68]. However, recent advances in understanding collective behavior of non–human organisms, flow of granular materials and crowd simulation have highlighted the need to examine the power–law relationships of time headways for empirical data on crowd movements at bottlenecks [26,27,33,58]. Therefore, inspired from the above studies, power–law fittings were performed for the entire datasets. Time headways distributions of egress crowds through bottlenecks displaying a power–law under different system conditions can be represented as shown in Equation 5.

$$P(\Delta t) \sim \Delta t^{-\alpha} \tag{5}$$

where α is known as the scaling parameter (or the exponent) which is a constant parameter of the power law distribution. Previous studies have used α value as the indicator of system flow and clogging condition [27,49,58]. Specifically, sometimes a system can be recognized as an unclogged state when $\alpha > 2$ plus a reasonable average flow, and a clogged state when $\alpha \leq 2$ plus a very low mean flow value as the increase of measuring time [27,58].

Under log–log scale coordinates, the tail of power–law distribution obeys a linear form as shown in Equation 6.

$$\ln P(\Delta t) = \alpha \ln \Delta t + C \tag{6}$$

We adopted the power–law fitting and goodness of fit test methods described in [69] into studying the time headway distributions. The scaling parameter α and lower–bounded $\Delta t_{min}$ were estimated, respectively (and the results were displayed in Fig. 5). Goodness–of–fit tests (Kolmogorov–Smirnov tests) were conducted and all the experiment data passed the tests. The results were in line with [26].

Since the power–law relations are valid only when $\Delta t \geq \Delta t_{min}$, the fitted power law tail could indicate the clogging time threshold if we regard $\Delta t_c$ (defined in [58] as the boundary indicator of flow and clogging, when $\Delta t \geq \Delta t_c$, we can consider the system as clogged) as $\Delta t_{min}$. Using clogging time threshold, we could describe the relationships between clog and power–law parameter $\Delta t_{min}$.

From Fig. 5 we could observe that in all scenarios, α values were never less than 2 which is in line with previous power–law fitting results for empirical data reported in [49,58] while $\alpha \leq 2$ existed in some granular cases [27]. Moreover, we could use the scaling parameter α to describe the degree of clogging. Under log–log scale coordinates, the greater α value was, the steeper the straight line fitted the power law tail was. In other word, the decay rates of probabilities for long clog rose, which reflected the better egress state. On the other hand, smaller α values indicated a relatively inefficient egress status as the probabilities for long clog decrease slowly. As can be observed from Fig. 5, with architectural configurations that performed better in terms of outflow (for e.g. corner exit) had higher α value as compared to the case which was not efficient in outflow (for e.g. middle exit).

Thus, with $\Delta t_{min}$ and α, clogging time and probabilities could be rigorously explored to identity the clogging events with different architectural configurations. This result with human data on clogging event is consistent with those obtained from egress data of non–human organisms [27,58] (i.e. any architectural configurations that enhance the outflow would have higher exponent α value compared to the other configuration that negates the outflow). It is therefore safe to assume that careful adjustments of architectural features as conducted in this study lessens the tendency of the system to clog.

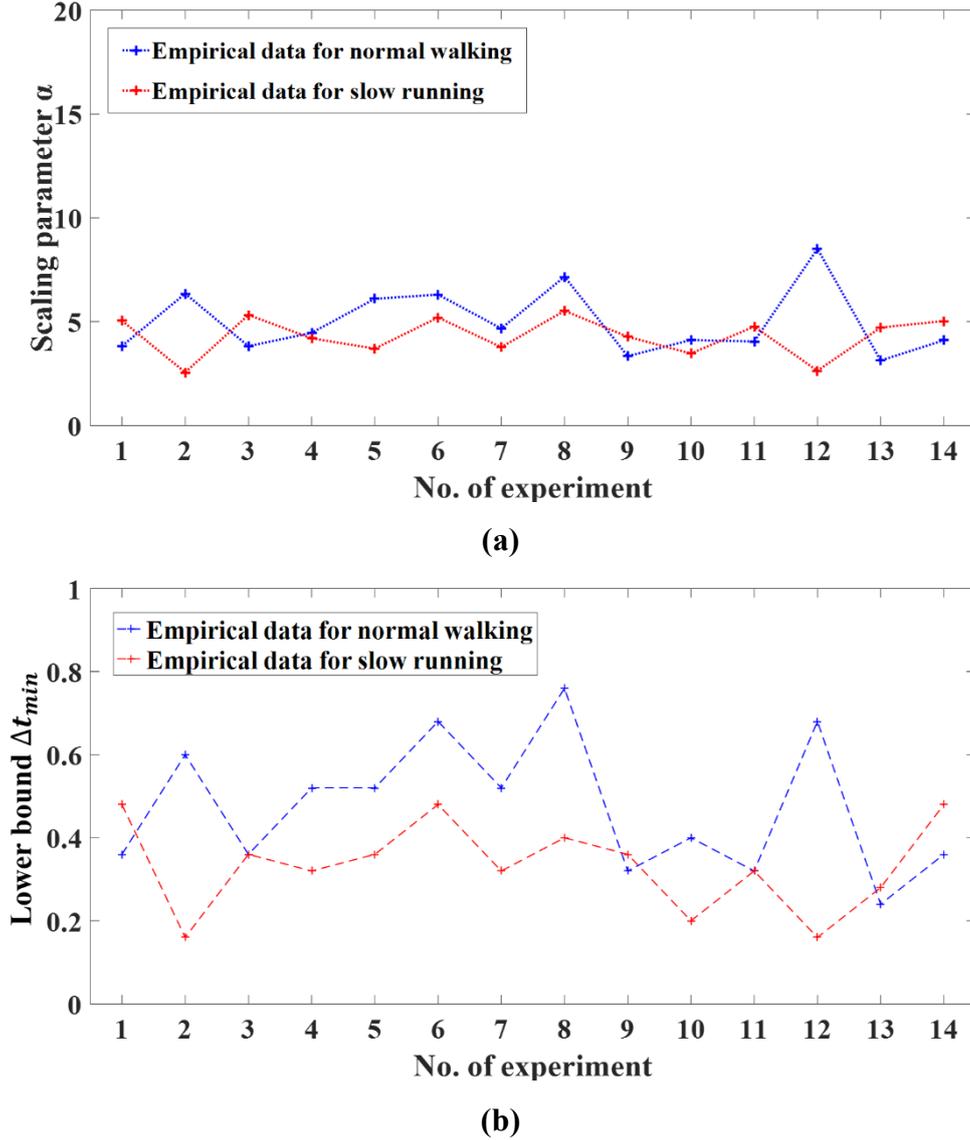

**Fig. 5.** Power–law fitting parameters distributions for different experimental setups (a): scaling parameter α; (b): lower bound $\Delta t_{min}$

## 5. Discussion and Conclusion

Effect of adjustment of structural features on the outflow of individuals at bottleneck has been of considerable interests to the research community on crowd dynamics. If seemingly small structural features of the physical environment can have disproportionate influence on the pedestrian escape dynamics, then there is enormous potential in the design of optimal layout of the escape area. This can have implications in the design of public space, major transport hubs and other public buildings for safe and efficient crowd movements. As such studies ranging from mathematical simulation to experiments with non–human organisms have been conducted in the past to examine this particular problem. However, comprehensive systematic study with complementary human data on the performance of different architectural adjustments at the bottleneck has been limited in the literature.

Inspired with the previous experiments with non–human organisms and simulation studies, we conducted a series of experiments to examine the effect of size and location of obstacle near an exit as well as the location of exit on the outflow of individuals. It was observed from the experiments that within a given escape area, careful adjustments of structural features can lead to substantial increase of the outflow of pedestrians. For example, with the architectural configuration that had corner exit with 60 cm diameter column located at 100 cm from the corner exit, the average specific flow was 3.18±0.34 Ped/s/m. However, in the case where the exit was located in the middle of the wall, the flow was 2.63±0.20 Ped/s/m. This result supported the position taken by some researchers based on prediction from simulation models and results from non–human organisms' experiments that there is a beneficial effect of architectural adjustments and obstacles near exit towards the outflow [2,4,17,32,56,70]. Further, our results also supported the position taken by some researchers who reported from their simulations that these obstacles are not always efficient in terms of outflow and that their performance depends on size and location of the obstacle [22,39,71]. For example, in contrast to the above case of corner exit, in the experimental configuration that had a 60 cm diameter column and located at 60 cm from the middle of the exit, the average specific flow was 2.57±0.09 Ped/s/m as compared to 2.57±0.09 Ped/s/m with middle exit free of obstacle. Thus the present evidenced based approach has provided insight into the differing decisions about the performance of these small architectural features of an escape area and how they contribute to the development of optimal design solutions for a given escape layout.

While the results from previous simulation studies and our present empirical study suggest that appropriately placed columns or pillars may provide benefits in terms of stabilizing the flow and increasing the throughput in case of high level of competition or motivation, some real–life observations by the researchers in the literature also point out the beneficial effect of columns in low level competition (normal conditions). For example, Helbing et al. [17,72] illustrates with a real–life observation of a pedestrian tunnel connecting two subways in Budapest's Metro System at Deak Tér where a series of columns in the tunnel stabilizes pedestrian lanes when the density of pedestrians and the variations of their desired speed are high. Hence, suitably placed and designed columns may serve both efficiency and safety functions for low and high level of competition or motivation. While columns at the exit may decrease the architectural value and depending on their design may reduce visibility at the exit, appropriate solutions can be developed in consultation with the architects. Some of the solutions proposed in the literature (for e.g. illuminated columns, telescope columns activated only in critical situations) are promising from practical implementation [17]. However, it is to be noted that obstacles in front of doors may have an effect on the perception of people on the functionality of the doors. Presently, to our knowledge all the studies have focused on the simulation studies, observational studies and the empirical study towards the understanding of the performance of obstacles in terms of crowd outflow and safety. However, it seems study focusing on people's perception of these obstacles near an exit /door and the functionality of doors is missing in the literature. There is a scope to integrate the expertise of engineers, physicists, architects and psychologists towards conducting systematic survey/experiments on people's perception on obstacles, functionality of doors and bottlenecks in the future.

A comparison of flow rate between normal walking and slow running conditions with our experimental setups demonstrated that the flow rate for slow running was always larger than normal walking. Hence, we could not observe FIS effect under slow running condition and instead noticed FIF effect in line with some other recent study [21]. We believe FIS effect could be more prevalent during strong competitive behavior and higher degree of threat and as such needs further scrutiny in future.

In terms of headway distribution, the present analysis provided support to the recent emerging approach of analyzing the power–law tail distribution of time lapses to identity the clogging events. It was demonstrated that the identification of clogging event is depended on the exponent α value. Moreover, higher exponent value is expected to the flows that are less prone to clogging as compared to the flows that are prone to clogging. This observation is in close agreement with recent study on clogging transition of biological and granular entities flowing through bottlenecks [27].

Thus, through our approach, it is possible to study a variety of scenarios at bottlenecks, potential problems, their consequences, and the outcome and effect of collective dynamics among different biological and physical systems. Further, the empirical data will provide foundation to verify and test simulation models intended to simulate pedestrian flow at bottlenecks. In future, it is recommended to examine other complex architectural configurations such as merging and/or diverging corridors and stairs. Also, there is need to further examine the physical and behavioral similarities and dissimilarities among the collective behavior of different biological entities and how they may help to develop design solutions that could enhance the crowd safety.

**Acknowledgements**

This research is sponsored by the transportation and technology transfer project of Jiangsu DOT (No. 2016Y21), the Scientific Research Foundation for the Returned Overseas Chinese Scholars of State Education Ministry. The third author would like to acknowledge the financial support from Victorian Department of Transport / Public Transport Victoria (PTV) (ARC Linkage project LP120200361). The authors would also like to thank the staff and students from Suqian College for their assistance in organizing and participating in the experiments. The authors also want to express thanks to Dr. Maik Boltes and his colleagues in Forschungszentrum Jülich GmbH for their help in the usage of PeTrack. Finally, we thank four anonymous reviewers whose constructive feedback has helped us to improve the paper.